\documentclass[12pt]{article}
\begin{document}

\title{Bohr's Conception of the Quantum Mechanical State of a System
and Its Role in the Framework of Complementarity}
\author{Henry J. Folse \\
Department of Philosophy\\Loyola University -- New Orleans\\
New Orleans LA, USA\\Email: folse@loyno.edu}
\date{17 June 2001}

\maketitle

\thispagestyle{empty}

\abstract{What Niels Bohr called the `epistemological lesson' of
`complementarity' was the result of reasoning analogically from the
classical conception of a mechanical state to a new quantum
mechanical conception of an `object' in a mechanical state. Bohr
proposed to redefine the `objectivity' essential for scientific
description in terms of the \textit{epistemological} demand for
unambiguously communicable descriptions of observational results, a
move which has profound consequences for how we can understand the
concept of the quantum mechanical state and the nature of the
`system' which is `in' this state.  Here it is argued that the old
notion of the `object' which is in a classical mechanical state is
drawn from a substance/property ontology derived from Aristotle's
analysis of categorical propositions.  In moving to describing a
system in a quantum mechanical state, the system that is `in' such a
state can no longer be so regarded as a \textit{substance}
possessing properties.  Bohr argues that the concept refers to an
\textit{interaction} which has a feature of wholeness or
`individuality' that implies that the distinction between `object
system' and `observing system' is relative to the context of the
description. This conclusion, in turn, implies the need for a
combination of complementary modes of description; however, because
of his reticence in making ontological claims, he failed to develop
this dimension of his new framework of complementarity.}

\section{The Conceptual Framework for the Description of Nature}

The concept of the \textit{quantum} mechanical state of a physical
system around which so much controversy swirls was formulated by
reasoning from an analogy to the \textit{classical} mechanical
state.  Niels Bohr repeatedly appealed to this kind of reasoning in
feeling his way from the `stationary states' of his original
quantized model of the atomic system of 1913 to his mature
understanding of the quantum formalisms of Heisenberg and
Schr\"odinger. As  a contribution to clarifying basic concepts, here
I propose to offer an analysis of how this crucial concept of the
`state' of a `physical system' was transformed from the `intuitive'
classical concept formulated in terms derived from ordinary human
experience into the context of the `mysterious' quantum theoretical
representation of the atomic domain.  This transformation, and the
general epistemological lesson it teaches, is at the heart of Niels
Bohr's thinking in formulating his new framework of complementarity.

In reconstructing the thought of an influential scientist one must
be careful not to impute to the historical actors a consciousness
which itself is the product of their very own contribution to the
scientific outlook, and hence how history often has come to perceive
them. Thus Newton was no Newtonian---certainly not what that name
would have meant even by the mid-eighteenth century.  And Einstein
had no conception of the Big Bang cosmology which is the boldest
fruit of his contribution to the scientific conception of the
universe.  In Bohr's case there are at least two ways which we can
say the product of his labors has disoriented the philosopher's
attempt to reconstruct how he saw the scientists' description of
nature.

The \textit{first} is the historical fact that Bohr came to play the
role of captain of the philosophically quite diverse crew of
Copenhagen.   Whatever it was that this assortment of geniuses who
professed loyalty to Bohr held in common was erected into something
like a creed under the banner of `The Copenhagen Interpretation,'
the confession of which---in the opening paragraphs of a
textbook---was held to expiate the author from demand for any further
philosophical explication.  `The Copenhagen Interpretation' was a
term Bohr never used and surely regretted.  Given its essentially
\textit{strategic} function in the history of atomic physics, a
sociological analysis of the party in the grip of the
\textit{Copehagenergeist}, such as that offered recently by
Beller,\cite{be} can provide an illuminating example of the dynamics
of scientific belief formation, but inevitably only from the point
of view of one who already knows where the story is going. Thus this
cannot establish how the actors in these events saw what it was they
were doing from \textit{their} side of the revolutionary divide.
Moreover even if we grant there is no single coherent account of
nature in the so-called `orthodox view'---regarded as a strategic
\textit{alliance} among distinct thinkers---that does not establish
that the \textit{individuals} associated with it do not have a
coherent view to call their own. But this does require that to make
out some coherent reconstruction of Bohr's thinking, we need to
penetrate through the Copenhagener's front lines, and look just at
this individual physicist as he approached these issues.

A \textit{second} way in which we lose the historical scent is that
we ask questions that arise from how we \textit{currently}
understand the quantum formalism, which of course was not an
available option in the midst of the quantum revolution.  Clearly
Bohr could not have thought in terms of mathematical representations
such as the collapse of the wave packet or operations on vectors in
Hilbert space---as we are wont to do today---for these came only
later after his creative work was done.  But even if we disregard its
ahistoricity, such an approach which moves from the mathematical
formalism to an account of nature, was the very opposite of Bohr's
philosophical \textit{modus operandi}.  His whole career was staked
on the building of atomic models and a concern with what the
physicist must do in order to make such models harmonize with the
empirical evidence.  The same approach has been proposed by Fuchs in
a letter to David Mermin, where he makes the following observation:

\begin{quote}
The issue in my mind is {\it not\/} to {\it start\/} with complex
Hilbert space, unitary evolution, the tensor product rule for
combining systems, the identification of measurements with Hermitian
operators, etc., etc., and {\it showing\/} that Bohr's point of view
is {\it consistent\/} with that.  Instead it is to start with Bohr's
point of view (or some variant thereof) and see that the {\it
precise\/} mathematical structure of quantum theory {\it must\/}
follow from it. Why complex instead of real? Why unitary, rather than
simply linear? Indeed, why linear?  Why tensor product instead of
tensor sum? And, so on. When we can answer {\it these\/} questions,
then we will really understand complementarity.\cite{fu}
\end{quote}

Although talk about the `quantum mechanical state' is central to
discussions which proceed from the current formalism, in fact that
phrase was never used by Bohr. If we want to recapture how Bohr
would have regarded this concept of a quantum mechanical state, we
must approach it from his side of the revolutionary divide, not from
that of the subsequent formalism.

By avoiding these disorienting ways of asking questions, we can
reconstruct a plausible account of Bohr's philosophy which is at
least coherent, but frustratingly incomplete.  To put things in
philosophical jargon, I would maintain that many have misread Bohr
as advocating an anti-realist interpretation of physics, whereas I
maintain that if we look at him in the midst of  the quantum
revolution, we see a realist trying to reconstruct the classical
conceptual scheme for the description of nature in order to deal
with a brute shock to the physicists' preconceptions about nature.
The result was what he called `complementarity' and what it
purported to be was a `new framework for the description of
nature.'  No matter what triumphs the Copenhagen Interpretation may
have had, Bohr's lifelong campaign to promote this new framework,
which he often called the `\textit{epistemological} lesson' taught
to us by atomic physics, was a failure.  Perhaps one reason is that
his `epistemological lesson' needed an \textit{ontological} lesson
to go with it, yet Bohr failed to provide any clear indication of
what such a revision would entail.  Here I would like to pick up the
trail and try to sketch out what I take to be the ontological
revisions suggested by the framework of complementarity.

Bohr's formation of his philosophy of science was more or less
contemporary with the positivism of the Vienna Circle, and he does
share much with the positivists, but his `big picture' in building
his philosophy of science really points at least as much towards
later post-positivist images of science associated with so-called
\textit{Weltanschauungen} analyses.  `Natural science' for Bohr is
that part of `human knowledge' which is concerned with the
`description of nature.'  A `conceptual framework' is necessary for
expressing the knowledge gained in this scientific description of
nature in an `objective' manner.  The scientist attempts to fit the
framework to agree with empirical evidence generated by the
extension of the empirical basis into new domains, such as the
interior of the atom.  This requires revision of the framework which
is constrained both by the demand for logical consistency and the
need to accommodate the recalcitrant experience.  Although
contemporary philosophers have been made very much aware of these
points by the works of Quine and Kuhn, it is worth pointing out that
Bohr was writing well before these thinkers had made such themes
familiar.\footnote{I suspect that this may be because all these
thinkers shared a common debt to pragmatism generally, and William
James in particular, but the evidence for the connection between
James and Bohr, I have to admit, is equivocal and it is not likely
it will ever be established to everyone's satisfaction.}

For Bohr the crux of the revision in our understanding of how this
conceptual framework attaches to nature lies in the need to redefine
the `objectivity' of the description of nature. In particular, the
epistemological lesson Bohr drew from this revolutionary transition
taught that the `objectivity' of a description cannot be based on
the ontological claim that the description refers to the properties
which define the mechanical state of a system isolated from
interaction with observers.  Hence he proposed to redefine the
`objectivity' necessary for scientific description in the epistemic
demand for unambiguously communicable descriptions of observational
results.  This move has profound consequences for how we can
understand the concept of the quantum mechanical state and the
nature of the `system' which is `in' this state.

\section{The Classical Mechanical State of the System}

The thesis I want to explore is that the old notion of the `entity'
or  the `object' which is in a classical mechanical state is derived
ultimately from a concept of `information' based on the Aristotelian
analysis of propositions and the substance/property ontology
associated with it.   Aristotle defined  `primary substances,'
\textit{i.e.}, the `entities' or `beings' of which the physical
world is composed, as that which is the ultimate subject of all
predicative assertions but which themselves can never be predicated
of anything else.  While such primary substances are the ultimate
individuals of the Aristotelian description of the physical
universe, all such assertions, \textit{i.e.} all statements of
`information,' predicate \textit{universals} of the
\textit{particular} substance which is its subject. These are the
substance's `accidents' or `properties.'

For Aristotle `scientific knowledge' arises only when the active
mind, \textit{nous}, grasps the set of universals which are
essential to that substance's being what it is; but in doing so,
knowledge can have as its object only the `whatness' of that
substance, never its individual `thisness,' its particularity.  The
Aristotelian solution was to make the substance into a hylomorphic
synthesis of form and matter with the universal form, the object of
scientific knowledge, incorporated into the `matter' which made a
substance into an individual particular being.  But while one can
speak intelligibly of \textit{formed} matter, the concept of
\textit{pure} `prime matter' stripped of all form, becomes---on this
analysis---ultimately unknowable, that of which no universal
predicate can be affirmed.  For this reason ultimately, in the
Aristotelian system, knowledge of the individual of this particular,
contingent world recedes from our grasp.

\textit{Ancient} epistemology was not concerned with the distinction
between the subject's perception---of which we are immediately aware
in consciousness---versus an `objective reality' which in some way
is imagined to lie `behind' that conscious experience. For Aristotle
we experience the world \textit{objectively} as it \textit{is}, and
the world \textit{is} as we objectively \textit{experience} it. But
in the \textit{modern} philosophy of the Enlightenment, the
mechanization of nature accompanying the Cartesian turn toward the
certainty of what is present in subjective consciousness led to a
conception of nature wholly unlike the world revealed in human
experience.  While the problem of accounting for knowledge of the
particular individual `thisness' of concrete beings confounded
Aristotelian ontology, the new mechanistic world-view could easily
deal with that problem by identifying the individual by its unique
\textit{spatio-temporal} locus.  Thus the `objects' of which the
physical universe is composed had to \textit{be} things of which
spatio-temporal locus, the property which makes a \textit{kinematic}
description possible, could be necessarily and universally
predicated.  Extension became the `essence' of physical substance.
This was significant for it meant that physical beings, the things
that were `in' classical mechanical states, must be the sorts of
things that can be \textit{pictured}; they are essentially, we might
say, `visualizable in space and time' even though (as Berkeley
pointed out) no one has ever seen extended substance
\textit{simpliciter}.

With Newton's achievement the mechanization of nature added to
extended substances the \textit{dynamical} properties necessary to
account for the \textit{causes of change} in kinematic properties
over time.  This then completed the list of properties objectively
possessed by `mechanical systems,' because together the kinematic
and dynamic properties were all the predicates needed for a complete
mechanical description of physical systems and their interactions.
Therefore to construct a mechanistic ontology one needed only this
small handful of properties which were solely those essential for a
mechanical description of extended substances.  Thus they came to
replace the bounteous supply of Aristotelian natural essences, as
the \textit{only} universals which could be truly predicated of
\textit{individual} beings as they \textit{objectively} exist in
Nature.  Aside from these true cases of objective predication, the
rest of those universals we affirm of the objects of sensory
experience have only \textit{subjective} validity; these properties
exist only in human sense perception of the world.

From the time of the scientific revolution initiating this classical
mechanical world-view, the concept of an `objective' description was
understood in this \textit{ontological} sense, as referring to a
description which was limited to predicating only those properties
actually possessed by the `object' being described.  These so-called
`primary qualities' of bodies are `real' in the sense of existing
apart from any observation of them; they are `out there' in the
`world.'  In contrast the far larger host of `secondary qualities'
which objects are perceived as having merely exist as the causal
effects in a \textit{subject's} consciousness of the mechanical
interaction between bodily sense organs and `external' bodies which
really possess only the primary qualities.  The mechanician's
conception of the physical universe may seem an austere shadow of
the luxuriant world of sensory experience, but it's the real thing.
It is what is required by an `objective' description.

Of course this distinction between primary and secondary qualities
arose in a Cartesian rationalist conception of the mechanical world
view, and once imported into empiricism by Locke, it was exposed as
an unfounded `philosophers' distinction by Berkeley. No empiricist
since Hume would countenance this rationalist hang up. In spite of
that fact, up to the quantum revolution this seventeenth century
understanding of the definition of the mechanical state of a
physical system as essentially the representation of the primary
qualities of a substance remained a standing rationalist bastion in
a field otherwise conquered by empiricism. Yet in creating its
concept of the `mechanical state' of the `system' from the older
ontology of `substance' and `accident,' and making this the basis of
`objectivity,' classical mechanism naively moved from a concept of
how we \textit{know} the world to how the world must \textit{be}
with a distinctly pre-modern ease.  Kant's revolution was
`Copernican' in that it showed you cannot simply `read off' the
structure of Being from the structure of Knowing, as the older
Enlightenment philosophers had naively presumed you could.  But
Kant's revolution notwithstanding, the continued triumph of
mechanism throughout the 18th and 19th centuries could easily be
used to support the most materialistic of ontologies based on a
univocal correlation of the `primary qualities' of the mechanical
description with the properties of  an objective `Reality.'

\section{Bohr's Rational Generalization of the
Concept of the `Mechanical State of the System'}

The central theme of Bohr's career was the effort to fashion a
successful theoretical model of the atomic system.  From his
experience with Rutherford at Manchester he entered the world of
physics with the conviction that classical mechanics could not be
successful at accounting for the stability of atoms by representing
them as classical mechanical systems of charged particles.  Although
the notion of electron orbits often assumes center stage in
presentations of Bohr's model, in fact Bohr saw the heart of his
`revolution' to be the postulate that atomic systems exist in a
quantized series of `stationary states' which defy the laws of
classical mechanics.  Interactions between atomic systems and the
electromagnetic field or other particles therefore imply
discontinuous changes of state.   ``Taking the indivisibility of the
quantum of action as a starting-point," Bohr tells us, he was led to
suggest ``that every change in the state of an atom should be
regarded as an individual process, incapable of more detailed
description, by which the atom goes over from one so-called
stationary state into another. According to this view, the spectra
of the elements do not give us immediate information about the
motions of the atomic parts, but each spectral line is associated
with a transition process between two stationary states....the
necessity of making an extensive use, nevertheless, of the classical
concepts, upon which depends ultimately the interpretation of all
experience, gave rise to the formulation of the so-called
correspondence principle which expresses our endeavors to utilize
all the classical concepts by giving them a suitable
quantum-theoretical re-interpretation." \cite{ATDN3} Thus \textit{ab
initio} Bohr's work was predicated on an outlook which demanded in
the atomic domain a new `\textit{quantum} theoretical
re-interpretation' of a physical system being in a `mechanical state'
to replace the classical mechanical conception.

Bohr's ground level argument appears in variations in many of his
essays. He starts with the observation that widening experience
often forces revisions of the conceptual scheme employed for the
description of nature.  He then turns to Planck's `discovery' (not
`invention') of the quantum of action as forcing such a revision.
Bohr calls this surprising discovery the `quantum postulate' and
maintains that it has been forced upon us by the need to account for
the known relations between matter and radiation.  Thus the quantum
revolution demands a revision of the conceptual framework for the
description of nature because at the atomic level we must describe
physical systems as changing their states \textit{discontinuously}.
Bohr expresses this point by saying that the quantum postulate
requires that `interactions' between `systems' be accorded a feature
of `wholeness' or `individuality.'

Right from his 1913 model of the atom---long before the development
of what we know as `quantum theory'---Bohr always emphasized two
aspects of his atomic model: \textit{First}, all information about
the properties of atomic systems comes through \textit{interactions}
between atomic systems and the field or other material systems in
which the atomic system changes from one `stationary state' to
another `stationary state' \textit{discontinuously}.
\textit{Second}, when no  dynamical exchange is taking place between
the atomic system and electromagnetic radiation or other material
particles, the atomic system exists in a `stationary state' isolated
from interaction. While one may construct familiar visualizable
spatio-temporal models of this atomic system in its `stationary
states' isolated from any interaction, such constructions are
\textit{purely theoretical abstractions} because by definition of
the `stationary state,' the properties of the `orbits' or otherwise
visualizable trajectories used in these kinematic `pictures' cannot
be observed, because observation \textit{is} interaction:
\begin{quote}
...any attempt to fix the space-time co-ordinates of the constituent
particles of an atom would ultimately involve an essentially
uncontrollable exchange of energy and momentum with the measuring
rods and clocks which prevents an unambiguous correlation of the
dynamical behaviour of the atomic particles before the observation
with their later behaviour. Inversely, every application of
conservation theorems, for instance to the energy balance in atomic
reactions, involves an essential renunciation as regards the
pursuance in space and time of the individual atomic particles. In
other words, the use of the idea of stationary states stands in a
mutually exclusive relationship to the applicability of space-time
pictures.\cite{Bohr32}
\end{quote}
The visualizable `space-time pictures' of the stationary states by
means of orbits in space and time cannot be understood realistically
because the energy exchanged between the atom and the field is a
function not of the orbital characteristics but of the differences
between the stationary states.  Because they involve such
discontinuous changes in state, interactions at the atomic level
cannot be described as mechanical processes taking place in space
and time.

Classical mechanics was able to get away with its conception of
defining the state of an isolated system by observation only because
the presumption of the continuity of interaction allowed one to make
the dynamical effect of the interaction involved in the observation
infinitesimally small.  In much the same way that it was commonly
held that relativistic mechanics is a `generalization' of the
`special case' of classical mechanics in which the speed of light is
imagined to be infinite, Bohr held that the quantum mechanical
description is a \textit{rational generalization} of the special
case of the classical description, where it is imagined the
parameter of action is defined for a \textit{continuous} range of
values.

Of course this view of progress which represents earlier theories as
`special cases' of later more general theories was common coin
during the heyday of positivism when Bohr wrote, but today's
philosophers are keenly aware of a well known argument against it by
Thomas Kuhn.\cite{Kuhn}  He argued that this alleged reduction of
earlier theories to special cases of later theories is misleading to
the extent that although the analogous theories are expressed in
parallel vocabularies, in the transition from one framework to the
next, crucial terms change meanings.  Although Bohr no where
advocates the radical semantic incommensurability which so
preoccupied philosophers in the decades following Kuhn's provocative
work, far from taking the standard positivist line, Bohr actually
anticipated Kuhn in cautioning that in this new quantum description
of atomic systems one must recognize that \textit{the classical
descriptive concepts employed attach to nature in a manner different
from that of their use in the classical context}. In particular, to
pass the criterion of `science' the revision of the `classical'
framework necessary to accommodate the quantum postulate must result
in a description of nature which is `objective,' but the
`objectivity' demanded of a description of the quantum state of a
system now has to be given a new meaning because of the semantic
shift to the new framework.

Let us see how Bohr reached this conclusion: What we would like to
determine by observation---based on our classically trained
expectations---is a space-time `picture' in which we `visualize' the
state of the system `objectively,' \textit{i.e.} how it is isolated
from any perturbation in its state resulting from the observation.
The system is imagined, along classical lines, as an entity which
possesses essentially those properties which define its classical
mechanical state.  But observations require an interaction with the
object system, which, because of the `indivisibility' of interaction
at the atomic level, alters its state in a non-determined way.  For
determining what state a system is in by means of observation, the
dynamical conservation principles must be employed to trace the
causal interaction between observed system and measuring
instruments.  Thus the classical description employed a
\textit{conjunction} of dynamical conservation principles for a
causal analysis of the observational interaction with a
spatio-temporal visualizable representation of the mechanical state
of the system.  But the indivisibility of interaction expressed by
the quantum postulate now requires that this conjunction becomes
transformed into a \textit{`complementary}' or \textit{`reciprocal}'
relationship, as expressed in Heisenberg's infamous relations.
Complementarity is thus first proposed as a point of view to replace
what Bohr called the `the frame of the ordinary causal description'
of the `ideal of causality' employed by the classical framework.  In
a nice summary from 1937 he writes:

\begin{quote}
However, a still further revision of the problem of observation has
since been made necessary by the discovery of the universal quantum
of action, which has taught us that the whole mode of description of
classical physics, including the theory of relativity, retains its
adequacy only as long as all quantities of action entering into the
description are large compared to Planck's  quantum. When this is
not the case, as in the region of atomic physics, there appear new
uniformities which cannot be fitted into the frame of the ordinary
causal description ....Indeed this circumstance presents us with a
situation concerning the analysis and synthesis of experience which
is entirely new in physics and forces us to replace the ideal of
causality by a more general viewpoint usually termed
`complementarity.' The apparently incompatible sorts of information
about the behavior of the object under examination which we get by
different experimental arrangements can clearly not be brought into
connection with each other in the usual way, but may, as equally
essential for an exhaustive account of all experience, be regarded
as `complementary' to each other.\cite{Bohr37}\end{quote} Note that
Bohr presents this line of reasoning entirely from  the quantum
postulate and his general philosophical position regarding the need
to revise conceptual frameworks with the expansion of experience
into new domains.  Nowhere does he appeal to the specific formalism
of Heisenberg or Schr\"odinger or the properties of a mathematical
representation or model.  While Bohr's formulation of this argument
was coeval with Heisenberg's derivation of the indeterminacy
relations, it was the outcome of a path on which Bohr had been set
long before 1927, rather than a move to explain or justify
Heisenberg's surprising mathematical derivation after the fact. Bohr
saw in the indeterminacy relations confirmation of a point of view
he had reached primarily from conceptual analysis.

Having established the need to revise the classical conceptual
framework, Bohr then calls attention to the fact that an objective
description of an observation requires an `object' to be
unambiguously distinguished from the physical systems employed as
`measuring instruments' in the observation.
\begin{quote}
This circumstance, at first sight paradoxical, finds its elucidation
in the recognition that in this region it is no longer possible
sharply to distinguish between the autonomous behavior of a physical
object and its inevitable interaction with other bodies serving as
measuring instruments, the direct consideration of which is excluded
by the very nature of the concept of observation in itself....In
particular, the frustration of every attempt to analyse more closely
the `individuality' of single atomic processes, symbolized by the
quantum of action, by a subdivision of their course, is explained by
the fact that each section in this course definable by a direct
observation would demand a measuring arrangement which would be
incompatible with the appearance of the uniformities
considered.\cite{Bohr37}
\end{quote}
By our free choice in designing a particular experiment we bring
about the occurrence of a particular `phenomenon,' as Bohr used that
word after EPR in 1935. What Bohr calls the `interpretation' of the
phenomenon as an `observation' or `measurement' which yields a
certain outcome requires describing the phenomenon as a dynamical
interaction between the `object' and the `measuring instruments.' To
make the distinction between object and observing instruments
\textit{unambiguous} in our description of the phenomenon, we must
employ theoretical notions taken from classical physics, but the
wholeness or `individuality' of that interaction required by the
quantum postulate implies that exactly where that distinction is
drawn is a function of the particular description and whatever it is
described as a measurement of. ``Ultimately, every observation can,
of course, be reduced to our sense perceptions," he tells us, but
the ``circumstance... that in interpreting observations use has
always to be made of theoretical notions entails that for every
particular case it is a question of convenience at which point the
concept of observation involving the quantum postulate with its
inherent `irrationality' is brought in....After all, the concept of
observation is in so far arbitrary as it depends upon which objects
are included in the system to be observed."\cite{Bohr27} It is this
arbitrariness inherent in the description of a indivisible
interaction as a measurement which later in response to EPR led Bohr
to contend that in describing the interaction as an observation
which determines the observed property of the object, our
description will be `ambiguous' unless the entire interaction---the
whole `phenomenon'---is described.  It is worth noting that the
`ambiguity' of which he speaks is not that it is `ambiguous' whether
the system is to be considered a `wave' or a `particle,' but rather
that it is the `description' of the state of the system which is
`ambiguous' because the experimental arrangements necessary to
determine both canonical state parameters are mutually exclusive. He
first introduces `complementarity' in his `Como paper' of 1927 with
this remark:
\begin{quote}
This situation has far-reaching consequences. On one hand, the
definition of the state of a physical system, as ordinarily
understood, claims the elimination of all external disturbances. But
in that case, according to the quantum postulate, any observation
will be impossible, and, above all, the concepts of space and time
lose their immediate sense. On the other hand, if in order to make
observation possible we permit certain interactions with suitable
agencies of measurement, not belonging to the system, an unambiguous
definition of the state of the system is naturally no longer
possible, and there can be no question of causality in the ordinary
sense of the word. The very nature of the quantum theory thus forces
us to regard the space-time co-ordination and the claim of
causality, the union of which characterizes the classical theories,
as complementary but exclusive features of the description,
symbolizing the idealization of observation and definition
respectively.\cite{Bohr27}\end{quote}

He observes, ``This situation has far-reaching consequences" but
aside from telling us that ``space-time co-ordination and the claim
of causality" are``complementary but exclusive features of the
description" it is far from clear what these consequences are.  Here
I want to suggest that these consequences are `far-reaching' because
they imply that \textit{the `objectivity' of the resulting
description can no longer be grounded in the ontological supposition
that the description of nature predicates properties of its `object'
which refer to properties actually possessed by the object system
apart from any observational interaction with it}.  Thus to save the
objectivity of the description, the `\textit{quantum} mechanical
state' of a physical system must be understood in a different sense
from the way the \textit{classical} mechanical state description was
understood.

In his new framework, Bohr moves `objectivity' from the
\textit{ontological} context given to it by classical mechanism and
reconstructs it on an \textit{epistemological} basis. `Objective'
description now means description which is unambiguously expressed
in terms of everyday human experience, of the `common framework
adapted to human use and daily life' such that everyone could agree
on it.  Again Bohr's reasoning precapitulates the subsequent
development of positivism.  In the 1920's and 30's when Bohr was
writing, positivists hoped to reduce the superstructure of
theoretical statements to foundational `incorrigible' observation
reports expressed in a purely observational vocabulary.  This would
have been the empiricists' surrogate for the rationalist doctrine of
primary properties.  They would refer directly to properties of
elements of pure experience, or sense data or whatever, unsullied by
any subjective `interpretation' put upon what is alleged to be
`directly given.'  But the quest for such a foundational vocabulary
was doomed.  Eventually positivists---as well as the pragmatists and
Popperian realists---had to retreat back to the conventionalist
position that there is no one privileged language or conceptual
scheme which attaches directly to empirical reality.  Alternative
possibilities exist, and the choice of the language that actually
does get chosen is made for pragmatic, rather than foundational,
reasons. This was Carnap's move to `physicalism': observation
sentences are simply those on which all rational persons with normal
sense organs will agree.  In this sense Bohr's new conception of
`objectivity'---like the positivist recourse to physicalism---
provides an empiricist foundational `observation' statement, but it
is only a relative foundation, relative to the particular framework
used to express this particular description of nature.  It is, I
think, fair to say that Bohr's proposed revision of the conception
of objective knowledge is at least as much `pragmatic' as
`positivistic,' but the point I want to make here is not its
philosophical pedigree, but that once `objectivity' is defined in
this manner, its basis is \textit{epistemic}, resting on what we are
justified in accepting or believing, rather than
\textit{ontological}, seated in the properties independently
possessed by an individual object, a body as described by classical
mechanism.

\section{The Ontological Lesson of Complementarity}

Because Bohr redefines `objectivity' in the manner he proposes, in
moving from describing systems in classical mechanical states to
describing them in quantum mechanical states the system which is
\textit{in} such a state can no longer be regarded as a
\textit{`substance possessing properties}' precisely because what we
predicate of the system in such a state is not, in general, regarded
as the `properties' possessed by a substance.   Now in the quantum
description, the `system' which is \textit{in} a quantum mechanical
state must be reconceived as an \textit{interaction} which has a
feature of `wholeness' or `individuality' that implies that the
distinction between `object system' and `observing system' necessary
for making a description of that interaction `unambiguous' is
\textit{relative} to the context of the description.     He
certainly recognized that this move would have ontological
consequences, for in the passage immediately preceding his inaugural
introduction of  `complementary' he tells us that ``...the quantum
postulate implies that any observation of atomic phenomena will
involve an interaction with the agency of observation not to be
neglected. Accordingly, an independent reality in the ordinary
physical sense can neither be ascribed to the phenomena nor to the
agencies of observation."\cite{Bohr27}   But aside from this merely
negative declaration that they do not have ``an independent reality
in the ordinary physical sense," Bohr leaves us in the dark
ontologically, when it comes to the sort of non-ordinary `reality'
we \textit{are} to ascribe to the objects of atomic physics. This
reticence in making ontological claims left this dimension of
complementarity undeveloped. So from here on we must extrapolate
from---or perhaps reconstruct---what Bohr has told us about his `new
viewpoint.'  Bohr developed an \textit{epistemological} lesson
because his argument terminated in a conclusion about how we gain
scientific \textit{knowledge} about physical systems at the atomic
level.  But epistemological claims about what kind of knowledge we
can or cannot have about such systems have \textit{ontological}
implications about what such systems which are in quantum mechanical
states must \textit{be}. Until such implications are understood, the
`foundations' of the quantum theoretical description of matter will
remain mysterious.

Traditional realists have aspired to construct ontological models on
the basis of a pattern of reasoning long enshrined in the classical
conception of an `objective' description within a substance/property
ontology.  Following such reasoning, a `realist' reading of
complementarity seems to produce paradoxes such as a particle which
traces a trajectory through the left slit, but `knows' whether the
right slit is opened or closed; or a cat in a superposition of alive
and dead states.  Such paradoxes do not arise because of any
application of the formalism, which in each case is self consistent
and in accord with the empirical data.  However, the paradoxical
character of the description lies in the disconnect between the
concept of \textit{being} the sort of `system' which could be in a
\textit{quantum} mechanical state, and our ontological
preconceptions about substances and their properties conditioned by
\textit{classical} mechanical expectations.

If we say that the `system' which is described as `in a quantum
mechanical state' is not understood via the traditional Aristotelian
conception of subject as \textit{substance}, that of which our
description truly predicates properties, but rather is an
indivisible phenomenon described as an \textit{interaction} of which
we predicate possible outcomes, the paradoxical quality of the
concept of the quantum mechanical state disappears.  This is what
Bohr called the unexpected feature of `atomicity' in the quantum
world.  What are indivisible are not tiny `bodies,' Cartesian
\textit{rei extensae}, but \textit{processes} or \textit{events} in
which the prior state of one system has a causal effect on the
subsequent state of another system.  Thus the ontological lesson
which the need for recourse to complementary descriptions teaches us
is that  the physical properties to which the classical concepts
refer now belong to the \textit{termini} of an indivisible
phenomenon; they define at one end its preparation state and at the
other its outcome state.  The quantum-mechanical `systems' of which
we predicate `quantum mechanical states'  are \textit{interactions}
with a `superposition' of probabilistically weighted possible
outcomes.

Much of our confusion, it would seem, arises from what amounts to an
equivocation on what is an `object' in the description of nature.
Classically the `object of our description,' a body possessing
primary kinematic and dynamic properties, was the \textit{same}
object which is the `system' that is `in a classical mechanical
state.'  Whether that system was described as `interacting' with
another system or as `isolated' from interaction classically was
\textit{irrelevant}. In the quantum theoretical description of
nature at the microlevel we can no longer move directly from
observed system to isolated system, but the tendency to do so is
deeply entrenched because the objects in the world of our ordinary
perceptual experience, which are the ultimate referents of any
description of an \textit{empirical} outcome, are objects which
\textit{can} be treated as `isolated' from any dynamical effects of
our perception of them. Consequently the predicative assertions we
make about \textit{them} can be treated as attributing properties to
substances in the manner long ago analyzed by Aristotle. When we
treat statements about the things described by quantum mechanics in
the same way, because of the quantum postulate, we are led to
paradoxical conclusions.  Therefore, the object of our description,
the `system' which is `in a quantum mechanical state,' cannot be a
substance possessing properties, but must be regarded as a whole
phenomenon which theory allows us to interpret as an
\textit{interaction} between measuring system and the object---the
`atomic system'---of which a measurement outcome is predicated. But
the quantum postulate, implying as it does the indivisibility of
such interactions, forces us recognize that the `object' in this
sense of that of which we predicate a measurement outcome is
\textit{not} the \textit{same} `object' which is the system `in a
quantum mechanical state.'

While I would contend that such an ontological lesson, once fully
absorbed into the physicist's conception of nature, would remove the
paradoxical or mysterious quality of the microdomain, there
remains---of course---something `surprising' about this quantum
description. It is that the objects of our description, these
interactions, do not have \textit{determined} outcomes.  If Bohr and
the \textit{Copenhageners} are right, \textit{indeterminacy is a real
aspect of Nature}; and quantum theory is, after all, the consequence
of a fundamental indeterminacy of nature at the microlevel.  This is
surprising against classical intuitions, but not `paradoxical' in
the same sense as a cat in a superposition of alive and dead states.

Of course we have had nearly a century to get used to the idea of
indeterminacy at the atomic level.  We're not so
\textit{classically} obsessed anymore, and that's not \textit{so}
terribly `surprising.'  But the ontological lesson---at least to a
realist---that remains also still `surprising' is that if the seeds
of being are not \textit{substances} possessing essentially spatial
locus but are \textit{interactions} which have a feature of
`indivisibility' or `atomicity,' then a single whole interaction may
very well have \textit{nonlocal} outcomes. Bohr's response to EPR
explicitly warned that a description would be `ambiguous'---and so
\textit{not} `objective' in his sense---if the entire phenomenon is
not included as a single indivisible whole in the description.
`Quantum systems' are `non-separable' precisely because on this view
they are not spatio-temporally located substances; we must, in
Bohr's words, `renounce space-time pictures.'  Thus in Bell type
phenomena spatially separated measurements on `particles' all form
part of the outcome of a single interaction in a single `quantum
mechanical state.'  Perhaps this remains surprising to us in a way
which I suspect it didn't seem to Bohr from his `new point of view
of complementarity' because we have not followed the same journey he
did.  After all, these violations of determinacy and locality had
their birth in Bohr's conception of the interaction between an atom
and the field as a `discontinuous' transition between
non-visualizable stationary states: in the famous words attributed
to Schr\"odinger: `those damned quantum jumps.'


\end{document}